\documentclass[12pt]{article}

\usepackage[paper=a4paper, margin=2cm]{geometry}
\usepackage[russian,english]{babel}
\usepackage[cp1251]{inputenc}
\usepackage{amsmath}

\usepackage[logos]{amsbib}

\begin{document}

\date{}
\renewcommand{\refname}{References}

\author{A.~Yu.~Samarin
}
\title{Deterministic nonlinear quantum evolution. Experimental verification.}

\maketitle

\centerline{Samara Technical State University, 443100 Samara, Russia}
\centerline{}

\abstract{\textsl{It is described a nonlocal interaction between entangled quantum objects, which is initiated by a process different from the reduction of the wave function. The scheme of an experiment realizing a deterministic nonlocal quantum evolution is proposed. In the case of the negative result of the experiment, the universal character of the integral wave equation with a kernel in the form of a path integral is questionable, otherwise faster-then-light communication is possible}. 



A local influence on a composite quantum system generates the wave function transformation in whole space simultaneously. Such a possibility was taken as a paradox contradicting the special relativity~\cite{bib:1}. However, the algorithm proposed by Bell for the  analysis of this paradox~\cite{bib:2} confirmed the existence of a specific nonlocal interaction between parts of entangled quantum systems~\cite{bib:3,bib:4,bib:5}. Since until now all the nonlocal effects were appeared in the form of nonlocal quantum correlations of probabilities, efforts to remove the contradiction were reduced to the proof of the impossibility of faster-than-light communication~\cite{bib:6,bib:7,bib:8}, in consequence of the  stochastic nature of the collapse. The corresponding theorem have been formulated~\cite{bib:9}. 

In order to overcome the difficulties associated with the stochastic form of single results of nonlinear evolution for realizing faster-then-light communication in\cite{bib:10}, it was suggested to use a nonlinear quantum evolution that differs from the von Neumann reduction and has a deterministic result. The possibility of such processes is the direct consequence of the integral wave equation with the kernel in the form of a path integral. The existence of these processes is obvious from the differential equation for reduced density matrix of an open system\cite{bib:11}. The indirect experimental evidence of this hypothesis is represented in\cite{bib:12}.

The basis for a nonlocal interaction is the possibility of a local effect on an open system, provided the measure of the wave function is conserved, which ensures simultaneous conversion of the wave function throughout space\footnote{This transformation cannot be reduced to the Sr\"odinger evolution and the corresponding time is usually negligible small compare with the time of the wave function formation determined by the characteristic time of motion along the virtual paths}. Entangled quantum systems are most suitable for such experiments because of their large size and convenience of changing the potential energy of system's individual  components. In order to increase the size of the composite system, we introduce into the quantum system a coherent photons ensemble.

Consider $ N $ identical particles (further simply atoms), each of which is in the same state of coherent superposition of four excited stationary states (see the figure).     
\begin{figure}
 \includegraphics[width=100mm,height=50mm]{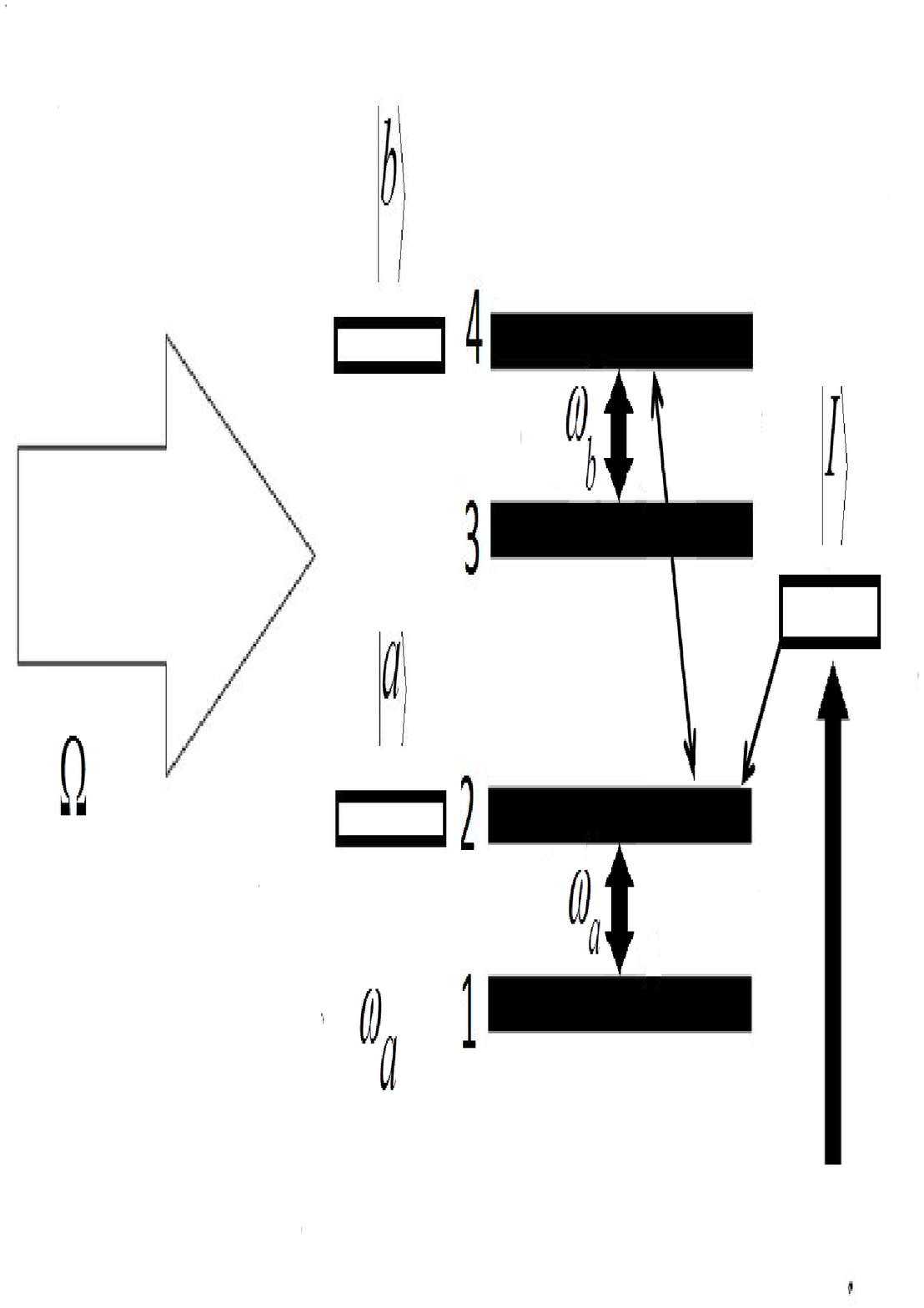}
\end{figure}
Atoms from the ground state are excited to the intermediate state $ | I \rangle $. Then a irreversible transition generates the state $ | 2 \rangle $. The coherent superposition $ c_2 | 2 \rangle + c_4 | 4 \rangle $ is created by the external coherent radiation. The system is inserted in the resonator having the resonator frequencies
 $ \omega_ {a} $ and $ \omega_ {b} $, which  coincide with the frequencies of atomic transitions $ \omega_ {21} $ and $ \omega_ {43} $. The interaction between the excited atoms and the resonator electromagnetic fields forms the coherent states $ c_1 | 1 \rangle + c_2 | 2\rangle + c_3 | 3 \rangle + c_4 | 4 \rangle $ (all decoherence processes are neglected) of the atoms. Since the photons of the resonator modes are emitted by the atoms of the system, they are in entangled states with the emitting atomic electrons. Indeed, the initial wave function of a single atom and corresponding electromagnetic field $ \Psi_ {t '} (q', a ') $ (where $ q' $, $ a '$ is, respectively, the variables of the excited atom and the vector potential amplitude of the electromagnetic field emitted by this atom at the initial time $ t '$) is transformed in accordance with the integral wave equation~\cite{bib:13}
\begin{alignat*}{2}
\Psi_{t}(q,a_{a},a_{b})=\int\int K_{t,t'}(q,q',a,a')\Psi_{t'}(q',a')\,dq'da'=\\=\int\int K_{22}(q,q',a,a')c_2(t')\Psi_2(q')\Phi_0(a')\,dq'da'+\\
+\int\int K_{21}(q,q',a,a')c_2(t')\Psi_2(q')\Phi_0(a')\,dq'da'+\\+\int\int K_{44}(q,q',a,a')c_4(t')\Psi_4(q')\Phi_0(a')\,dq'da'+\\
+\int\int K_{43}(q,q',a,a')c_4(t')\Psi_4(q')\Phi_0(a')\,dq'da',
\end{alignat*} 
where $ c_1 (t ') = c_3 (t') = 0 $; the transitions corresponding to $ K_ {22} $ and $ K_ {44} $ do not change the state of the system; the amplitudes $ K_ {21} $ and $ K_ {43} $ are determined by the path integrals~\cite {bib:14} in the forms
\begin{equation*}
 K_{21}=\int\int \Biggl(\exp\frac{i}{\hbar}\bigl(S_{p}[q(t)]+S_f[a_{a}(t)]+S_{int}[a_{a}(t),q(t)]\bigr)\Biggr)[dq(t)][da_a(t)] ,
\end{equation*}
\begin{equation*}
 K_{43}=\int\int \Biggl(\exp\frac{i}{\hbar}\bigl(S_{p}[q(t)]+S_f[a_{a}(t)]+S_{int}[a_{b}(t),q(t)]\bigr)\Biggr)[dq(t)][da_b(t)], 
\end{equation*}
where $ S_ {p} [q (t)] $, $ S_ {f} [a (t)] $, $ S_ {int} [q (t), a (t)] $ are the actions of the electron, field and interaction of the electron with the field, respectively; the vector potential amplitudes subscripts indicate the resonator mode. The actions $ S_ {int} [q (t), a (t)] $  depends both on the electron and on the field variables, which does not allow us to factorize the corresponding transitions amplitudes. Thus we have the entangled state of the electrons and the electromagnetic fields.

Let us neglect the radiation losses. Then for the wave function of the entangled state of an excited electron and the electromagnetic field radiated by this electron, we have
\begin{equation}
\Psi^{j}(q^{j},a_{a}^{j},a_{b}^{j},t)=\sum\limits_{l=1}^{4}c^j_{l}(t)\Psi_{l}(q^{j},t)\Phi_{al}(a_{a}^{j},t)\Phi_{bl}(a_{b}^{j},t).\label{eq:math:ex1}
\end{equation}
The index $ j $ identifies the atomic electron and the electromagnetic field radiated by it, the index $ l $--- the stationary state of the atom. The wave functions of the field oscillators in the \eqref {eq:math:ex1} correspond to the electronic states of the atoms
\[ \Phi_{al}=\left\{
\begin{array}{@{\,}r@{\quad}l@{}}
\Phi_{a}^{0} & \text{if } l=2,l=3,l=4. \\ \Phi_{a}^{1} & \text{if } l=1,  \end{array}\right. \]
\[ \Phi_{bl}=\left\{
\begin{array}{@{\,}r@{\quad}l@{}}
\Phi_{b}^{0} & \text{if } l=1,l=2,l=4, \\ \Phi_{b}^{1} & \text{if } l=3\\  \end{array}\right. \]
where $ \Phi_ {a}^{0} $, $ \Phi_{a}^{1} $ are the wave functions of the vacuum and single-photon states of the oscillator $ a $; $ \Phi_ {b}^{0} $, $ \Phi_ {b}^{1} $ are the analogous wave functions of the oscillator $ b $. The wave function of all electron-photon pairs that do not interact with each other will have the form
\begin{equation}
\Psi(q^{1}...q^{N},a_{a}^{1},...,a_{a}^{N},a_{b}^{1},...,a_{b}^{N},t)=\prod\limits_{j=1}^{N}\Psi^{j}(q^{j},a_{a}^{j},a_{b}^{j},t).\label{eq:math:ex2}
\end{equation}
The norm of the wave function~\eqref{eq:math:ex2} is determined by the coefficients $c ^ j_l $ in ~\eqref{eq:math:ex1}. If the atoms are uniformly excited in space, then the moduli of the expansion coefficients are the same for all atoms, and for the normalization expression, we have
\begin{alignat*}{2}
\int\dots\int\biggl|\Psi_{t}(q^{1}...q^{N}a_{a}^{1},...,a_{a}^{N},a_{b}^{1},...,a_{b}^{N})\biggr|^{2}\,dq^{1}\cdot\cdot\cdot \,dq^{N}\,da_{a}^{1}\cdot\cdot\cdot\,da_{a}^{N}\,da_{b}^{1}\cdot\cdot\cdot \,da_{b}^{N}=\\=\Biggl(\sum_{l=1}^{4}|c_{l}(t)|^{2}\Biggr)^{N}=
\sum_{\substack{n_{1},n_{2},n_{3},n_{4}\\n_{1}+n_{2}+n_{3}+n_{4}=N\\n_{1},n_{2},n_{3},n_{4}>0}}\frac{N!}{n_{1}!n_{2}!n_{1}!n_{2}!}|c_{1}|^{2n_{1}}|c_{2}|^{2n_{2}}|c_{3}|^{2n_{3}}|c_{4}|^{2n_{4}}=1,
\end{alignat*}
where $ n_1 $, $ n_2 $, $ n_3 $, $ n_4 $ are the occupation numbers of stationary states. Using Stirling's approximation and introducing the notations $ X_ {l} = \frac{n_{l}}{N}$, for the probabilities of the states having different occupation numbers, we obtain
\begin{equation}\label{eq:math:ex3}
P(n_{1},n_{2},n_{3},n_{4})\sim\Biggr(\biggl(\frac{|c_{1}|^{2}}{X_{1}}\biggr)^{X_{1}}\biggl(\frac{|c_{2}|^{2}}{X_{2}}\biggr)^{X_{2}}\biggl(\frac{|c_{3}|^{2}}{X_{3}}\biggr)^{X_{3}}\biggl(\frac{|c_{4}|^{2}}{X_{4}}\biggr)^{X_{4}}\Biggr)^{N}.
\end{equation}
In the case of a macroscopic number of atoms of the system, this dependence is not zero in an exceptionally narrow range of occupation numbers, in the vicinity of values that correspond to the maximum probability. Thus, the occupation numbers are almost exactly determined by the moduli of the coefficients of the expansion of the wave function of a single atom.

Let us consider a single atom located in the radiation field of other system atoms. The expansion coefficients in ~\eqref{eq:math:ex1} can be obtained using the perturbation theory in integral form~\cite{bib:13}. The unperturbed action for the virtual path of an electron in an atom is assumed to be independent on the resonator field. Then the electromagnetic field is considered as a minor term in the action expansion 
\begin{equation*}
S_{int}^{j}=\int\sum_{\substack{k=1\\k\neq j}}^{N} U_{int}^{jk}dt=\frac{e}{c}\int\limits_{t'}^{t}{\bf A}^j\,{\bf\dot{q}}^{j}\,dt,
\end{equation*}
where the indices $ j $, $ k $ identify the atoms of the ensemble; $ S_ {int} ^ {j} $ is the perturbing part of the action functional for the path of the particle $ j $;  $ U_ {int} ^ {jk} $ is the interaction energy between the excited electrons of the atoms $ j $ and $k$; ${\bf A} ^ j $  is the value of the vector potential of the radiation field in the region of the location of the atom $ j $. For perturbation theory it is necessary that $ S ^ {int} << \hbar $. Corresponding choice of the time interval can provide this condition. For the expansion coefficients we have
\begin{equation*}
c_{n}(t)=\sum_{m=1}^{4}\lambda_{nm}c_{m}(t'),
\end{equation*}
where the first order transition amplitude  is determined by the expression
\begin{equation*}
\lambda_{nm}=\delta_{nm}exp\frac{i}{\hbar}E_{m}(t-t')+\lambda^{(1)}_{nm}.
\end{equation*}
The first order correction $\lambda^{(1)}_{nm}$ has the form~\cite{bib:13}:
\begin{eqnarray*}
\lambda_{nm}^{(1)j}=-\frac{i}{\hbar}\int\limits_{t'}^{t}\biggl(\int\psi^{*}_{m}\sum_{k} U^{jk}\psi_{n}dq\biggr)exp\frac{i}{\hbar}\Bigl( E_{n}(t''-t)-E_{m}(t''-t')\Bigl)\,dt''=\\
=-\frac{i}{\hbar}\int U^{j}_{nm}exp\frac{i}{\hbar}\Bigl( E_{n}(t''-t)-E_{m}(t''-t')\Bigl)\,dt'',
\end{eqnarray*}
where $U^{j}_{nm}=\int\limits_{t'}^{t}\psi^{*j}_{m}\sum_{k} U^{jk}\psi_{n}^jdq$ is the matrix element of the interaction energy of the atom $j$ with other atoms of the system. Then for the first corrections to the amplitudes of the transitions we have
\begin{equation*}
|\lambda_{12}^{(1)j}|=|\lambda_{21}^{(1)j}|=\frac{1}{\hbar}\bigl|U_{12}^{j}\bigr|(t-t')=\frac{1}{\hbar c}\bigl|a_{a}J_{12}\bigr|(t-t'),
\end{equation*}
\begin{equation*}
|\lambda_{34}^{(1)j}|=|\lambda_{43}^{(1)j}|=\frac{1}{\hbar}\bigl|U_{34}^{j}\bigr|(t-t')=\frac{1}{\hbar c}\bigl|a_{b}J_{34}\bigr|(t-t').
\end{equation*}
The matrix elements $J_{12}$ and $J_{34}$ characterize the properties of the transitions in atoms with respect to the interaction with radiation in the resonator; the  vector potential amplitudes $ a_{a} $, $ a_{b} $ is considered as quantities averaged over the wavelength of the radiation, which results in the corresponding averaging of the occupation numbers. For the time dependence of the absolute value of the expansion coefficients, we obtain
\[ \left\{\begin{aligned} &|c_{1}^{j}(t)|^{2}=|c_{1}^{j}(t')|^{2}+|\lambda_{12}^{(1)j}|^{2}|c_{2}(t')|^{2},\\
&|c_{2}^{j}(t)|^{2}=|c_{2}^{j}(t')|^{2}+|\lambda_{21}^{(1)j}|^{2}|c_{1}(t')|^{2},\\
&|c_{3}^{j}(t)|^{2}=|c_{3}^{j}(t')|^{2}+|\lambda_{34}^{(1)j}|^{2}|c_{4}(t')|^{2},\\
&|c_{4}^{j}(t)|^{2}=|c_{4}^{j}(t')|^{2}+|\lambda_{43}^{(1)j}|^{2}|c_{3}(t')|^{2}.
\end{aligned}\right. \]
The first and last equations show that the absolute values of the states occupation numbers take on a stationary values
\begin{alignat}{2}\label{eq:math:ex44}
|c_{1}(t)|^{2}=|c_{2}(t)|^{2}\nonumber\\
|c_{3}(t)|^{2}=|c_{4}(t)|^{2}.
\end{alignat}
Then, from the normalization of the wave function, we obtain
\begin{equation*}
|c_{1}|^{2}+|c_{2}|^{2}+|c_{3}|^{2}+|c_{4}|^{2}=1
\end{equation*}
Taking into account~\eqref{eq:math:ex3}, we have
\begin{equation}\label{eq:math:ex55}
\frac{|c_{a}|^{2}}{|c_{b}|^{2}}=\frac{|\lambda^{(1)}_{12}|^{2}}{|\lambda^{(1)}_{34}|^{2}}=\frac{|U_{a}|^{2}}{|U_{b}|^{2}}=
\frac{|a_{a}J_{12}|^{2}}{|a_{b}J_{34}|^{2}}=\frac{n_{1}}{n_{3}},
\end{equation}
where $ U_{a} $, $ U_{b} $ are the matrix elements of the interaction energy of an atom with the field oscillators $ a $ and $ b $; $ c_ {a} = c_ {1} = c_ {2} $, $c_{b}=c_{3}=c_{4}$. From ~\eqref{eq:math:ex1},~\eqref{eq:math:ex44},~\eqref{eq:math:ex55} we obtain
\begin{alignat}{2}\label{eq:math:ex6}
\prod\limits_{j=1}^{N}\Biggl(\frac{c_{1}}{\sqrt{2}}\biggl(\Psi_{1}^{j}\Phi_{a}^{1}+\Psi_{2}^{j}\Phi_{a}^{0}\biggr)\Phi_{b}^{0}+\frac{c_{3}}{\sqrt{2}}\biggl(\Psi_{3}^{j}\Phi_{b}^{1}+\Psi_{4}^{j}\Phi_{b}^{0}\biggr)\Phi_{a}^{0}\Biggr)=\nonumber\\
=\prod\limits_{j=1}^{N}\biggl(c_{1}\Psi_{a}^{j}+c_{3}\Psi_{b}^{j}\biggr),
\end{alignat}
where the states $ \Psi_{a}^{j} $ and $ \Psi_{b}^{j} $ are the wave functions of the stationary entangled states of the $ j $ -th atom and the electromagnetic field emitted by it.

The ratio of the occupation numbers $ \frac{n_{1}}{n{3}} $ is determined by the excitation conditions of the atoms and can be initially arbitrary. On the other hand, in accordance with~\eqref{eq:math:ex55} it is in a single-valued correspondence with the ratio $\frac{|a_{a}|}{|a_{b}|}$. Then, if the external influence changes the value $ \frac{n_{1}}{n{3}} $  in a local volume of space, then the ratio $\frac{|a_{a}|}{|a_{b}|}$ changes and, therefore, the value $ \frac{n_{1}}{n{3}} $ changes simultaneously in whole space. If the intensity and the rate of increase in the external influence is quite high, then for a sufficiently large spatial size of the system, we can register the response to the external action faster than velocity of light.

Let the quantum oscillator pulse having the frequency $\omega_{\Omega}=\omega_{24}$ affect on the macroscopic subsystem $A$. Let $M$ be the number of atoms in the subsystem. Suppose that the intensity of this pulse is high to consider $n_{a}=n_{b}$ for the subsystem $A$. Then for atoms in the rest space (that are not affected by the quantum oscillator pulse), we have
\begin{equation}\label{eq:math:ex9}
\frac{n_{a}}{n_{b}}=\frac{|c_{a}|^{2}}{|c_{b}|^{2}}=\frac{\bigl|U_{a}^{j}\bigl|^2}{\bigl|U_{b}^{j}\bigl|^2}=\Biggl|\frac{a^p_aJ_{12}}{a^p_bJ_{34}}\,\frac{(N-M)\frac{n_{a}^{(0)}}{N}+\frac{1}{2}\,M}{(N-M)\frac{n_{b}^{(0)}}{N}+\frac{1}{2}\,M}\Biggl|^2,
\end{equation}
where $a^p_a $, $ a^p_b $ are the amplitudes of the vector potentials of single photons. The expression~\eqref{eq:math:ex9} determines the limit ratio of the occupation numbers of states of the subsystem $ B $ after the external influence. In general, it differs from the initial one.

This experiment is not only important from point of view of faster-then-light communication. In the case of the negative result of the experiment, the universal character of the integral wave equation is questionable. In this case, the evolution of the wave function due to a change in the weight of subsets of virtual paths~\cite{bib:11} is absent, and  the integral evolution law becomes equivalent to the Schr\"odinger equation describing linear local processes. A positive result indicates the presence of nonlocal nonlinear quantum processes, other than the reduction of the wave function in the measurement. In this case, many processes in open quantum systems, such as the nonlocal interaction of entangled quantum objects, transitions between stationary states of quantum systems under external action, decoherence, collapse of the wave function, etc., represent a class of the quantum phenomena united by the common evolution mechanism, different from the Schrodinger evolution. In addition, the deterministic nature of nonlinear quantum evolution attaches an epistemic nature of quantum probability~\cite{bib:10}.

\vfill\eject


\begin{thebibliography}{10}

\Bibitem{bib:1}
\paper Can quantum mechanics description of physical reality be considered complete?
\by A. Einstein~A., B. Podolsky~B., N. Rosen~N.
\jour Physical Review
\yr 1935
\vol 47
\pages 777--780
\url{http://link.aps.org/doi/10.1103/PhysRev.47.777}

\Bibitem{bib:2}
\paper On the Einstein-Podolsky-Rosen paradox
\by Bell~J.\,S.
\jour Physics
\yr 1964
\vol 1
\pages 195--200

\Bibitem{bib:3}
\paper Proposed Experiment to Test Local Hidden-Variable Theories
\by Clauser~J.\,F., Horne~M.\,A., Shimony~A., Holt~R.\,A.
\jour Phys. Rev. Lett.
\yr 1969
\vol 23
\pages 880--883
\url{http://dx.doi.org/10.1103/PhysRevLett.23.880}

\Bibitem{bib:4}
\paper Test of local hidden-variable theories
\by Clauser~J.\,F., Freedman~S.\,J.
\jour Phys. Rev. Lett.
\yr 1972
\vol 28
\pages 938--941
\url{http://link.aps.org/doi/10.1103/PhysRevLett.28.938}

\Bibitem{bib:5}
\by A. Aspect
\paper Bell's inequality test: more ideal than ever
\jour Nature
\yr 1999
\vol 398
\pages 189--190
\url{http://dx.doi.org/10.1038/18296}

\Bibitem{bib:6}
\paper Bell's theorem and the different concepts of locality
\by Eberhard~P.\,H.
\jour Nuovo Cimento B
\yr 1978
\vol 46
\pages 392--419
\url{http://link.springer.com/article/10.1007%2FBF02728628#page-1}

\bibitem{bib:7}
\paper A general argument against superluminal transmission trought the quantum mechanical measurement process
\by  Ghirardi~G.\,C., Rimini~A., Weber~T.
\jour Lett Nuovo Cimento
\yr 1980
\vol 27
\issue 10
\pages 293--298
\url{http://link.springer.com/article/10.1007%2FBF02817189#page-1}


\bibitem{bib:8}
\paper Quantum mechanics and faster-than-light communication: Methodological considerations
\by Ghirardi~G.\,C., Weber~T.
\jour Nuovo Cimento B
\yr 1983
\vol 78
\issue 1
\pages 9--20
\url{http://link.springer.com/article/10.1007%2FBF02721378#page-1}

\Bibitem{bib:9}
\paper Quantum information and relativity theory
\by Peres~A., Terno~D.
\jour :	Rev. Mod. Phys.
\vol 76
\pages 93--123
\yr 2004
\url{https://arxiv.org/abs/quant-ph/0212023}
\url{http://journals.aps.org/rmp/abstract/10.1103/RevModPhys.76.93}

\Bibitem{bib:10}
\by A.Yu.Samarin
\paper Can quantum objects be point-like particles?
\arxiv \href{https://arxiv.org/abs/1710.10154} [quant-ph]

\Bibitem{bib:11}
\by  A.Yu.Samarin
\paper  Nonlinear dynamics of open quantum systems
\arxiv \href{https://arxiv.org/abs/1706.09405v3} [quant-ph]

\Bibitem{bib:12}
\paper Quantum mechanics vs relativity: an experimental test of the structure of spacetime
\by Emelyanov~S.~A.   
\jour :	Phys. Scr. 
\vol  T151
\papernumber 014012
\yr 2012
\url{iopscience.iop.org/1402-4896/2012/T151/014012}
\url{https://arxiv.org/abs/0901.0088v22}

\Bibitem{bib:13}
\book Quantum Mechanics and Path Integrals
\by R.~P.~Feynman, A.~R.~Hibbs
\publaddr New York
\serial  International Earth \& Planetary Sciences
\publ McGraw-Hill Co.
\yr 1965
\crossref{http://www.amazon.com/Quantum-Mechanics-Integrals-Richard-Feynman/dp/0070206503}

\Bibitem{bib:14}
\book Path Integrals in Quantum Mechanics
\by J.~Zinn-Justin
\publ Oxford Press
\yr 2004
\publaddr Oxford
\crossref{http://dx.doi.org/10.1093/acprof:oso/9780198566748.001.0001}

\end{thebibliography}
\end{document}